\documentclass[a4paper,10pt]{article}

\usepackage[utf8]{inputenc}
\usepackage{fancyhdr}
\usepackage{authblk}
\usepackage[a4paper]{geometry}

\usepackage{authblk}

\usepackage{caption}
\usepackage{cite}
\usepackage{amsmath,amssymb,amsfonts}
\usepackage{algorithmic}
\usepackage{graphicx}
\usepackage{textcomp}
\usepackage{xcolor}
\usepackage{tabularx}

\usepackage[
	colorlinks=true,
	linkcolor=black,
	citecolor=black,
	urlcolor=blue
]{hyperref}
\usepackage{orcidlink}

\usepackage[final]{changes}

\def\BibTeX{{\rm B\kern-.05em{\sc i\kern-.025em b}\kern-.08em
    T\kern-.1667em\lower.7ex\hbox{E}\kern-.125emX}}

\usepackage[nolist]{acronym}

\begin{acronym}
	\acro{BESS}{Battery Energy Storage System}
    \acro{CHP}{Combined Heat and Power}
	\acro{DT}{Digital Twin}
	\acro{HEMS}{Home Energy Management System}
	\acro{EV}{Electric Vehicle}
	\acro{MiD}{Mobility in Germany}
	\acro{PV}{photovoltaic}
    \acro{RMSE}{Root Mean Square Error}
	\acro{SCR}{Self-Consumption Rate}
	\acro{SOC}{State of Charge}
	\acro{SSR}{Self-Sufficiency Rate}
	\acro{SPF}{Seasonal Performance Factor}
\end{acronym}

\newcommand{\PV}{\ac{PV} }

\fancypagestyle{firstpagefooter}{%
	\fancyhf{}
	
	\fancyfoot[L]{Accepted version of the paper submitted to IEEE SusTech, 2026\\
		DOI: \href{https://dx.doi.org/10.1109/sustech67720.2026.11536190}{10.1109/sustech67720.2026.11536190}\\979-8-3315-9258-5/26/\$31.00 \copyright~2026 IEEE}
}

\title{Quantification and Surrogate Model Estimation of Spatial-Temporal Carbon Intensity Factors}

\author[1]{Daniel R. Bayer\footnote{Email address: daniel.bayer@uni-wuerzburg.de}~~\orcidlink{0000-0002-4063-4097}}
\author[1]{Jonas Schiller\footnote{Email address: jonas.schiller@uni-wuerzburg.de}~~\orcidlink{0009-0007-8483-2339}}
\author[1]{Thanh Mai Pham\footnote{Email address: thanh-mai.pham@stud-mail.uni-wuerzburg.de}}
\author[1]{Marco Pruckner\footnote{Email address: marco.pruckner@uni-wuerzburg.de}~~\orcidlink{0009-0000-7241-119X}}
\affil[1]{Modeling and Simulation Lab, University of W\"urzburg, W\"urzburg, Germany}
\date{}

\begin{document}

\maketitle

\begin{abstract}
  The transition toward sustainable urban infrastructure is driving a rapid increase in distributed energy resources, particularly among prosumers equipped with solar \PV installations. Such \PV systems are highly dependent on solar radiation, which fluctuates over time and across different city districts, leading to substantial spatial and temporal variations in local carbon intensity factors. 
  The carbon intensity factor is a critical parameter for optimizing energy consumption and reducing emissions in context of smart building control, electric vehicle charging or district heating systems. 
  Hence, high-resolution information on local carbon intensity factors is crucial for intelligent control strategies that minimize environmental impact across urban infrastructures.

  In this paper, we introduce a novel approach for calculating spatially and temporally resolved carbon intensity factors on an urban district level.
  The results show that, even within a single city, these factors range from~0 up to 316~g/kWh for an exemplary summer noon.
  We further develop a transferable surrogate model that estimates these factors using only a limited set of input parameters, typically available to municipalities or grid operators, aligned with census grid cell information. For the surrogate model we compare both traditional decision tree based methods and a state of the art neural network architecture.
  Our findings demonstrate that the surrogate model can provide highly accurate estimations, particularly in densely built urban areas, supporting data-driven sustainable city operations and indicating transferability even to regions with limited available information.
  A case study further reveals that neglecting local variations can lead to emission estimation errors of up to 9\% for an exemplary building.
\end{abstract}

\thispagestyle{firstpagefooter}

\acresetall

\section{Introduction}
In 2024 alone, more than 450~GW of solar \PV were installed around the world, with a large share being installed on rooftops of residential and commercial buildings, transforming them into so-called prosumers that both produce and consume electricity~\cite{irena2025}. This trend contributes to lower carbon emissions, enhances local energy self-sufficiency, and reduces reliance on energy imports, all while offering economic benefits for building owners and occupants~\cite{2021_Lau}. Additionally, the electrification of other sectors, particularly heating and transportation, represents a key strategy for achieving substantial emission reductions in urban environments~\cite{Knobloch.2020}. Here, residential \PV installations in combination with heat pumps or \acp{EV} present potential to further reduce carbon emissions of individual households~\cite{2024_Bayer_a_HeatPumpRetrofitAnalysis}.
However, both the carbon savings and the achievable self-sufficiency rates are highly dependent on the specific location, ranging from 28\% in Poland~\cite{2020_Jurasz_PV_SSR_SCR_Poland} up to 46\% in Germany~\cite{2024_Bayer_d_LocalImpactOfFutureEVs_DTs} and 50\% in Italy~\cite{2021_Ciocia_PV_SSR_SCR_Italy}.
Similarly, the carbon intensity factors of electricity supplied to buildings varies greatly across different regions and countries, driven by their specific electricity generation mix, which in turn are influenced by climatic conditions~\cite{ember2025}.
Therefore, the environmental benefits of local solar \PV generation can vary between locations, highlighting the need for spatially and temporally resolved carbon intensity factors to arrive on better decisions about \ac{EV} charging and energy optimization.

In this context, tools such as \cite{DATASET_FFE_CO2Mix} and \cite{ELECTRICITYMAPS} provide real-time data on carbon emission factors at the national or regional electricity grid level, typically based on the reported generation mix. 
Based on the current emission situation, control decisions for load shifting of large consumers like \ac{EV} charging stations or heat pumps can be made by \ac{HEMS} to reduce the carbon footprint of buildings \cite{2023_Sen_Distributed_MPC_EMS}.
Nevertheless, \ac{PV} installation potential and actual deployment can vary considerably across districts within a single city~\cite{2020_Jurasz_PV_SSR_SCR_Poland}.
For example, suburban areas with smaller buildings typically offer better conditions for PV deployment than densely built-up city centers with high-rise buildings~\cite{2016_Gagnon_PVPotential}.
Thus, the actual local carbon intensity factors can be expected to differ highly even within a single city, which has a direct impact on the quality of the control decisions made by \ac{HEMS}.
In a broader perspective, this also implicates that the national carbon intensity factors may substantially differ from the actual local factors.

Whereas the temporal changes of national-level carbon emission factors have been considered in multiple publications \cite{LINDBERG2022,2016_Wang_CarbonIntensity} and tools \cite{DATASET_FFE_CO2Mix,ELECTRICITYMAPS}, the intra-city differences have received little attention.
To address this gap, this work (i) systematically quantifies intra-city variability using high-resolution smart meter data and different generation technologies like \ac{PV}, wind power and \ac{CHP}, and (ii) develops a transferable surrogate model capable of reproducing these local factors from a minimal set of input features typically available to municipalities or grid operators.
The results of (i) demonstrates that fine-grained modeling is not only possible but also practically relevant, as spatial deviations in carbon intensity can reach magnitudes that directly influence local control and planning decisions.

The rest of the paper is structured as follows: Related work is presented in Sec.~\ref{sec:relwork} and the method for computing the cell carbon intensities is presented in the first part of Sec.~\ref{sec:method}.
Thereupon, the estimation model for the carbon intensity factors is described.
The results of the two presented methods, i.e., the carbon intensity computation and succeeding the surrogate model performance, are presented in Sec~\ref{sec:results}.
The results are discussed in Sec.~\ref{sec:discussion} and the paper concludes in Sec.~\ref{sec:conclusion}.

\section{Related Work}
\label{sec:relwork}
Understanding and estimating the spatial distribution of carbon emissions is a critical step toward effective climate policy and urban energy planning. Numerous studies have explored emissions from various sources such as transportation, buildings, and industry, using approaches ranging from bottom-up activity-based models to top-down atmospheric inversion techniques \cite{2017Gately, Moran_2018}. 
Moreover, \cite{2025_Li_CarbonFootprintChina} highlight the importance of refining carbon footprint accounting methods for power systems by incorporating spatial and temporal variability.
However, localized grid emission intensity has received less attention with the first investigations published in the recent years. 
For instance, \cite{LINDBERG2022} analyze locational differences in marginal carbon emission intensity at the transmission grid level by examining how load shifts affect the solution of the DC optimal power flow.
On a more granular level, \cite{2024_Aryai_CarbonEmissionsWithHighResolution} develop a method to model real-time carbon emission intensity at high spatial resolution down to the substation level. Their findings from Australia highlight that carbon intensity can vary significantly even within a single city, with urban centers often showing higher values than surrounding rural areas. Similarly, \cite{2025_Sugano_SpatioTemporalCarbonIntensity} propose a method to assess intra-city variations in carbon intensity using smart meter data. In contrast to \cite{2024_Aryai_CarbonEmissionsWithHighResolution}, their approach accounts for hourly behind-the-meter \PV self-consumption and grid electricity intensity, revealing substantial spatial heterogeneity linked to the local use of \PV installations. While these recent publications mark important progress, they also come with certain limitations. For instance, the approach by Sugano et al.~\cite{2025_Sugano_SpatioTemporalCarbonIntensity} requires detailed smart meter time series for all consumers and only considers \ac{PV} as local generation technology, ignoring wind power or \ac{CHP}. Although the adoption of smart meters is increasing worldwide, access to such data remains limited due to privacy and security concerns~\cite{2017_Asghar_SMDPrivacyConcerns}.

In summary, existing methods for estimating local carbon intensity are either highly data-intensive or limited in the range of generation technologies they consider.
Moreover, they offer limited guidance on how to generalize results to areas with sparse or no high-resolution consumption data.
To address these gaps, we present a novel two-step approach: first, we use detailed smart meter data to derive local carbon intensity factors accounting for diverse generation technologies; and second, we develop a machine learning-based surrogate model that enables spatial generalization based on easily accessible regional characteristics.

\section{Method}
\label{sec:method}

\subsection{Considered region and used data}
\begin{figure}[htbp]
    \centerline{\includegraphics[width=0.75\textwidth]{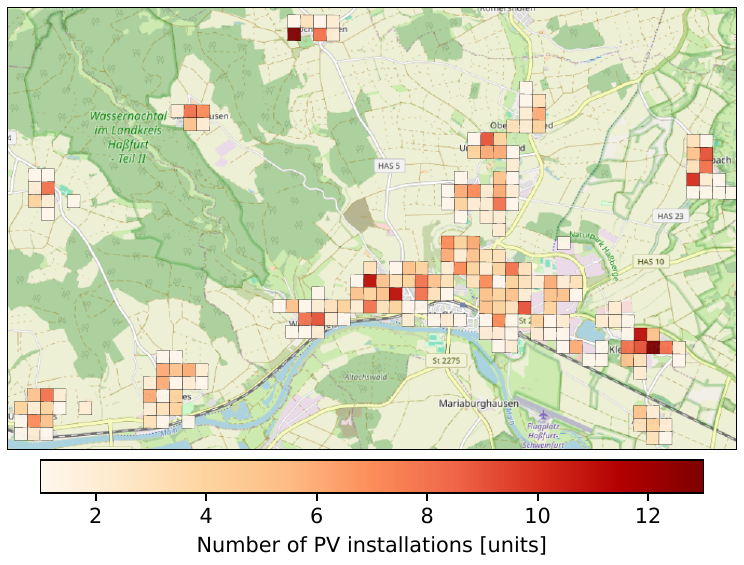}}
    \caption{\PV installation per grid cell for the town of Ha{\ss}furt, Germany. Map copyright by OpenStreetMap contributors 2025.}
    \label{fig:pv_installations_per_cell}
\end{figure}

We consider a town with around 16{,}300 inhabitants in Germany.
This town was selected, because it exemplifies a very common settlement structure in Germany, characterized by a primary city center and multiple smaller suburban districts \cite{2018_Krehl_TopologyOfUrbanCenters}.
A notable feature is the non-continuous spatial development between the core urban area and these suburbs, often interspersed with agricultural land, forests, or designated green spaces.
This structure contrasts with the often more expansive and continuously developed urban cultivation observed in other parts of the world, such as North America \cite{2018_Krehl_TopologyOfUrbanCenters}. Figure~\ref{fig:pv_installations_per_cell} provides a visual representation of this town structure highlighting the separated suburbs of the town, alongside the density of \PV installations within each analysis grid cell.
For the spatial discretization required by our analysis, we utilized grid cells measuring 200m~x~200m, consistent with the standard grid system employed by the German census \cite{LAEA_GeoGitter_Germany}.

From the local utility company, we obtained the smart meter data (demand and feed-in) with an hourly resolution for almost all buildings in the supply area, including 4{,}602~residential and around 1{,}000~non-residential buildings for~2021.
In this year, the town had over 700 solar \PV installations with a total peak power of 14~MWp (separated into rooftop and open-space \PV installations).
Moreover, 255~\PV installations are individually metered, i.e., there is an additional meter for the buildings' electricity consumption.
Additionally to the smart meter data, we obtained meta data on the metered appliances, especially if there is a \PV installation or a \ac{CHP} installed.
\added{During a preprocessing step, the smart meter time series were screened for completeness and plausibility.
Missing or invalid hourly readings were imputed for gaps of up to 500 hours per meter and year.
Only meters exceeding this threshold were excluded.}

\subsection{Estimation of the PV power per existing installation}
\label{sec:method_PV_power_estimation}
For the latter analysis of local carbon intensity factors, it is essential to separate the \PV generation from the residential electricity consumption, which are typically recorded together in most smart meter datasets.
One challenge in this separation is the lack of direct information on the rated peak power of the connected \PV installations in the majority of cases.
However, for a subset of buildings within our dataset, the installed \PV capacity (in $kW_p$) is known, provided by the utility company.
This subset of known installations provides a basis for estimating the peak power for the remaining buildings where this information is missing.

\begin{figure}[htbp]
    \centerline{\includegraphics[width=0.7\textwidth]{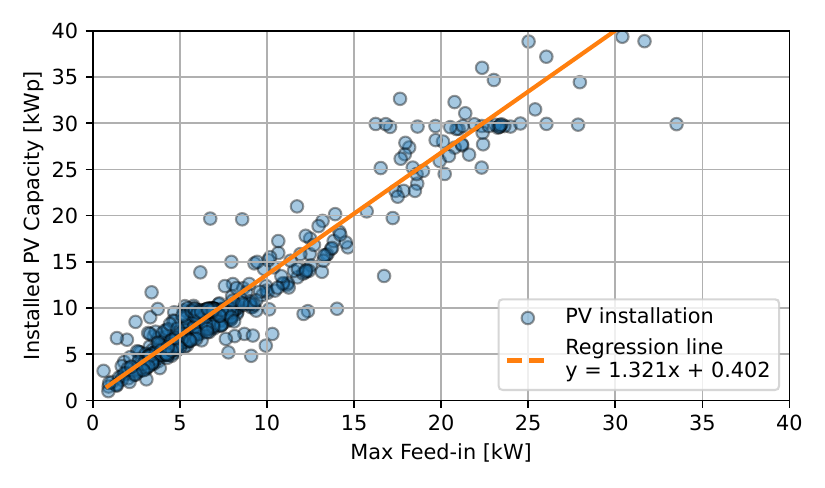}}
    \caption{Scatter plot of maximum annual metered feed-in and installed \PV power known by the utility operator. The regression result is depicted as orange line.}
    \label{fig:pv_capacity_estimation}
\end{figure}

The estimation methodology is based on the idea that the maximum electricity feed-in recorded by the smart meter over the period of a year is strongly correlated with the installed \PV capacity.
Therefore, we employed a linear regression model trained on the buildings with known capacities.
This model predicts the installed \PV capacity (in $kW_p$) as a function of the maximum observed annual feed-in power derived from the smart meter time series.
As visualized in Fig.~\ref{fig:pv_capacity_estimation}, this approach yields a high degree of fitting fidelity, achieving an $R^2$ value of 0.917.
The high correlation gives evidence for the idea of the estimation method.
This regression-based estimation technique aligns with similar methodologies employed in the literature, such as \cite{2022_Liu_PVPowerEstimationUsingSMD} which also utilize peak values in smart meter data.
Although more complex \ac{PV} power estimation or load disaggregation methods based on deep learning exist \cite{2024_Zhang_PVLoadDisaggregationUsingDL}, we use a linear model due to its simplicity and high accuracy.

\subsection{Calculation of local carbon intensity factor}
\label{sec:method_carbon_intensity_factor_computation}
The method for computing spatially resolved carbon intensity factors at the grid cell level is inspired by Sugano et al.~\cite{2025_Sugano_SpatioTemporalCarbonIntensity}, adding more generation technologies with different carbon intensities and creating a fulfillment schema.
First of all, all available smart meters (for electricity consumption and generation) are grouped according to their geographical location, assigning each meter to its corresponding census grid cell.
The overall objective is to determine the aggregated energy generation and consumption within each grid cell using this smart meter data.
While this is straightforward for buildings without local generation, a common challenge arises with \ac{PV}-equipped buildings, as many smart meters record only the net exchange with the grid (local consumption minus local \ac{PV} generation) rather than the gross consumption and generation separately.

\begin{figure}[htbp]
    \centerline{\includegraphics[width=0.8\textwidth]{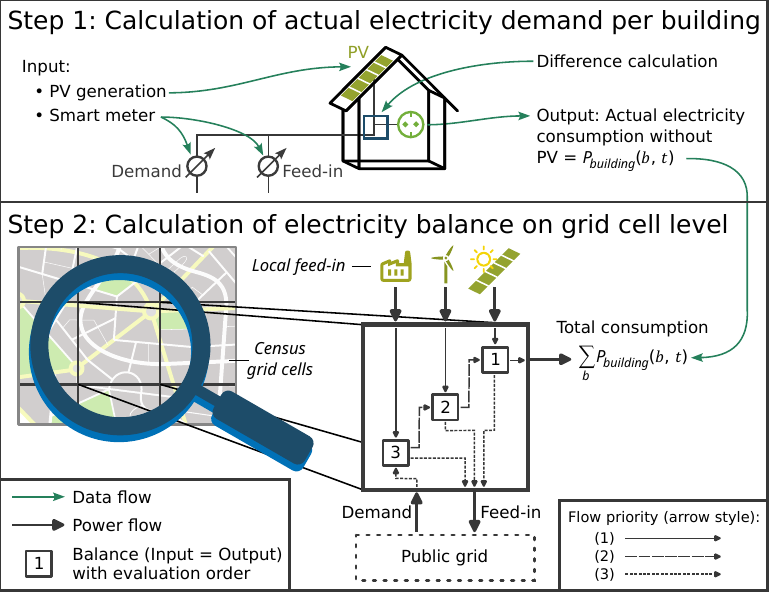}}
    \caption{Overview of the applied method for calculating the local carbon intensity factors.}
    \label{fig:calculation_method_overview}
\end{figure}

To overcome this limitation, we estimate the actual \PV generation for each relevant installation using a simulation-based approach, distinguishing our method from the statistical techniques used in \cite{2025_Sugano_SpatioTemporalCarbonIntensity}.
Subsequently, for each meter associated with a \PV system, the estimated original hourly electricity demand $P_{building}(b,t)$ of building $b$ at each hourly time step $t$ is reconstructed by combining the net meter readings with the simulated \ac{PV} generation $P_{PV}(b,t)$ for building $b$, as shown in \eqref{eq:raw_balance_computation},
\begin{equation}
    \delta(b,t) = P^{+}_{meter}(b,t) - P^{-}_{meter}(b,t) + P_{PV}(b,t)
\end{equation}
\begin{equation}
    \label{eq:raw_balance_computation}
    P_{building}(b,t) = \max\{ \delta(b,t), \; 0 \}
\end{equation}
where $P^{+}_{meter}(b,t)$ represents the metered grid demand, and $P^{-}_{meter}(b,t)$ denotes the power exported into the grid (metered feed-in) of building $b$ at time step $t$, see also step 1 in Fig.~\ref{fig:calculation_method_overview}.
To avoid negative values for the power of the building, we set $P_{building}(t)$ to zero if it would otherwise be negative.

The required \ac{PV} generation time series, $P_{PV}(b,t)$, are computed using the PVLib simulation library \cite{pvlib}, driven by local solar radiation data for the considered town. The necessary peak power parameter for each installation is obtained either from known meta data delivered by the utility company or is estimated using the method as described in Sec.~\ref{sec:method_PV_power_estimation}.
Since reconstructing the actual buildings' demand might alter the perceived annual consumption compared to raw meter data, a calibration step for the \ac{PV} model is performed as described in Sec.~\ref{sec:results_calibration}.

With the gross building demand ($P_{building}(b,t)$ for \ac{PV}-equipped buildings, or simply $P^{+}_{meter}(b,t)$ for others) and local generation sources identified, we compute the emissions on the level of every cell.
In this paper, we consider \ac{PV} installations, wind installations and \ac{CHP} plants as electricity generation technologies.
As these different technologies come along with different carbon intensities, we compute the local feed-in sum per technology per cell.
For \PV and wind power, we set the carbon intensity to zero, as both technologies do not have operational emissions.
For the \ac{CHP} plants, carbon emissions of 307~g/kWh are assumed, which is the mean value reported by \cite{2020_Siddiqui_EmissionsGHGBiogasPlants}.
These factors, combined with the aggregated energy flows within each cell, allow for the determination of the local carbon intensity, as detailed subsequently.

In this second step, we aggregate the computed time series for the actual building electricity demand $P_{building}$, the simulated \PV feed-in $P_{PV}$ and the metered feed-in for wind power and \ac{CHP} for all meters within a specific grid cell.
Thereupon, we determine the energy balance within each cell by calculating the self-consumption of locally generated electricity.
This situation is depicted in step 2 of Fig.~\ref{fig:calculation_method_overview}.
We assume a prioritization order where demand is first met by local \PV generation, followed by wind generation, and then \ac{CHP} generation, which corresponds to the ascending sorting according to the carbon emissions per component.
Prioritizing especially local \PV seems plausible as this rooftop \PV generation is typically directly consumed in the building.
Any remaining demand after exhausting local generation is covered by importing electricity from the external grid.
For this external grid, emissions are assumed to align with the national electricity mix.
The time series used in this study for the carbon intensity of the national electricity mix is derived from a public dataset \cite{DATASET_FFE_CO2Mix} that computes the national average by considering the national-wide sum of generation by technology per hourly time step multiplied by the technology-specific emissions.
The total emissions $C(c,t)$ per grid cell $c$ and time step $t$ are computed by multiplying the self-consumed amount $\varphi_g(c,t)$ per generation technology $g\in \{PV, Wind, CHP\}$ with the specific carbon intensity factor $\mu_g$ of this technology $g$ and adding emissions caused by possibly remaining grid demand,
\begin{equation}
    C(c,t) = \left( \sum_{g\in G} \varphi_g(c,t) \cdot \mu_g \right) + \varphi_{grid}(c,t) \cdot \mu_{grid}(t)
\end{equation}
where $G = \{PV, Wind, CHP\}$ denotes the set of considered generation technologies, $\varphi_{grid}(c,t)$ denotes the remaining grid demand to cell $c$ at time step $t$ and $\mu_{grid}(t)$ denotes the grid emissions per~kWh at time step $t$.
The value of $\varphi_g(c,t)$ can be computed iteratively as stated above.
Finally, the local carbon intensity factor $I(c,t)$ per cell $c$ and time step $t$ is obtained by dividing the hourly cell emissions by the total hourly electricity demand, i.e.,
\begin{equation}
    I(c,t) = C(c,t) \cdot \left( \sum_{b \in \mathrm{B}(c)} P_{building}(b,t) \right)^{-1}
\end{equation}
where $\mathrm{B}(c)$ denotes the set of buildings in cell $c$.
For time steps where the total demand is negligible (e.g., below 1 kW), the local carbon intensity is set to zero to ensure numerical stability.

\subsection{Estimation model for local carbon intensity factors}
Although the roll out of smart meter data is accelerating, a complete coverage will not be reached in the foreseeable future. Therefore, we investigate what data is required to calculate the hourly carbon intensity of city regions on the grid cell level and develop a surrogate model with predictive qualities under a limited amount of information. 
In the first step towards our surrogate model, we first reduce the description of a cell to the following parameters:
\begin{itemize}
    \item Number of \ac{PV} Installations
    \item Sum of Peak Power over all Installations
    \item Number of Wind Power Installations
    \item Number of \ac{CHP} Installations
    \item Number of \ac{BESS} Installations
    \item Number of Buildings
    \item Number of Residential Buildings
    \item Total Yearly Electricity Demand
    \item Total Yearly Electricity Feed-In
\end{itemize}
Local grid operators and municipalities typically have access to each of the aforementioned attributes within their jurisdiction. Additionally, grid operators often possess data on either the yearly net demand for each electricity consumer or the yearly total feed-in of prosumers, which is crucial for financial accounting. The presence of unregistered \ac{PV} installations can be estimated through manual counts or satellite imagery~\cite{malof2015}.

As the characteristics of city regions can vary quite drastically, we cluster the cells based on their attributes using kmeans~\cite{macqueen1967}. We obtain the optimal number of clusters based on the elbow method, e.g. identifying the point at which the average distance to the cluster centroid does not decrease significantly further when adding new clusters. Through this method, we identify 9 clusters. Reducing the number of clusters leads to decrease in average prediction quality as specific characteristics are missed by the prediction algorithms. Table~\ref{tab:cluster} gives the number of cells per cluster as well as the their characteristics such as average number of buildings and total installed solar \ac{PV} peak power.
\begin{table*}[htbp]
    \caption{Characteristics of Clusters, all values are the average over the clusters total values}
    \centering
    {
    \footnotesize
    \begin{tabular}{ll|c|c|c|c|c|c|c|c|c}
       \multicolumn{2}{l|}{Cluster}  &  1 & 2 & 3 & 4 & 5 & 6 & 7 & 8 & 9 \\ \hline
       \multicolumn{2}{l|}{Number of Cells} & 68 & 7 & 1 & 111 & 23 & 1 & 1 & 45 & 1 \\
       \multicolumn{2}{l|}{\textit{Average values for}}  & & & & & & & & & \\
 \hspace{9pt} & \ac{PV} size (kWp)& 42.1 & 40.2 & 869.9 & 22.6 & 40.7 & 2023.7 & 1229.3 & 57.24 & 68 \\
       & \ac{BESS} Capacity (kWh) & 4.6 & 1 & 2000 & 1.33 & 3.1 & 0 & 0 & 6.3 & 0 \\
       & N. of Buildings & 62 & 264.7 & 4 & 18.2 & 186 & 60 & 325 & 115.35 & 646 \\ 
       & N. of Residential Buildings & 24.1 & 144.1 & 0 & 5.2 & 83.2 & 23 & 192 & 40.6 & 154
    \end{tabular}
	}
    \label{tab:cluster}
\end{table*}
Clusters 3,6,7 and 9 are outliers with either a very large battery (3) or a high total installed PV peak power (kWp) (6) or are densely populated (7,9). Cluster 1 represents cells with on average 62 buildings and 42 kWp while Cluster 4 represents more sparsely populated cells with only 18 buildings. Cluster 4 describes the most populated cells apart from the two in outliers in 7 and 9, while cluster 2 and 5 describe cells with higher density but low \PV installation rates.

Besides the time-independent characteristics of each cell, the model also requires time-varying, exogenous inputs to estimate carbon intensity on an hourly basis. These include solar radiation, which varies both temporally and spatially, and the grid emission level, which changes over time. Additionally, we perform extensive feature engineering on the data, considering factors such as time of year, day of the week, and scaling adjustments. 
A sample therefore consists of the cell characteristics as well as the current situation and the model aims to predict the cells specific carbon intensity at this timestep. We note that this setup, together with existing forecasts on irradiation and grid emission intensity level could also provide a location specific emission forecast. In this work, however, we focus on achieving a good transferability to previously unseen regions, which is why for each cluster with more than one cell we train the models on 80~\% of the cells in  a cluster and use the other 20~\% for validation. 

As surrogate models, we test two decision tree based models, a classic decision tree and eXtreme Gradient Boosting (XGBoost), which is an ensemble learning method based on decision trees and gradient boosting~\cite{2016_Chen_XGBoost}. Additionally, we investigate the ability of an advanced Neural Network Architecture called TabNet, which has previously shown promising results in regression tasks~\cite{arik2021tabnet}. TabNet is a deep learning architecture specifically designed for tabular data, combining the interpretability of decision trees with the learning capabilities of neural networks. It utilizes a sequential attention mechanism to select meaningful features at each decision step, enabling dynamic feature selection and efficient learning. These properties make it a good candidate to outperform classic decision trees on regression tasks. The hyperparameter of each model are carefully optimized to ensure robust and reliable performance. To assess model performance, we use the coefficient of determination ($R^2$) and the Relative Root Mean Squared Error (RRMSE), defined as follows:
\begin{equation}
\text{RRMSE} = \frac{1}{\bar{y}} \sqrt{\frac{1}{n} \sum_{i=1}^{n} (y_i - \hat{y}_i)^2}
\end{equation}
where $y_i$ denotes the true values, $\hat{y}_i$ the predicted values, $\bar{y}$ the mean of the true values, and $n$ the number of samples. The $R^2$ score provides a measure of how much of the variability can be explained through the model while the RRMSE gives a good indication of the performance error.

\section{Results}
\label{sec:results}

\subsection{Calibration of the PV generation model}
\label{sec:results_calibration}
To ensure the reliability of the \PV generation model, the initial annual simulated \PV generation calculated using PVlib is compared against empirical feed-in time series derived from 255~smart meters monitoring exclusively PV installations within the study area.
The comparison reveals a consistent overestimation of the \PV generation model for over 98\% of these individually metered installations.
Shading \cite{2017_Tripathi_EffectsOfPVShading} and other effects like cell aging or soiling of the modules \cite{2023_Rahman_PV_Degradation} reduces the actual generation in the real world.
As both are not directly considered in the PVlib, overestimation is plausible and needs to be corrected through calibration.

To determine an appropriate calibration factor, the total annual simulated \PV generation summed over all 255 installations was compared to the total annual metered \PV feed-in from these same installations.
A calibration factor ($k_{\text{calib}}$) was calculated as the ratio of the total metered energy to the total simulated energy.
The total annual metered feed-in was 3.42~GWh, while the total simulated generation was 5.13~GWh.
Thus, the calibration factor was calculated as $k_{\text{calib}} = \frac{3.42~GWh}{5.13~GWh} \approx 0.67$.
This calculation yielded a calibration factor of $k_{\text{calib}} = 0.67$.
Applying this factor, effectively reducing the raw PVLib simulation output by 33\%, ensures that the total calibrated annual simulated \PV generation matches the total annual metered \PV generation when summed across all monitored installations.

\begin{figure*}[h!]
    \centerline{\includegraphics[width=\textwidth]{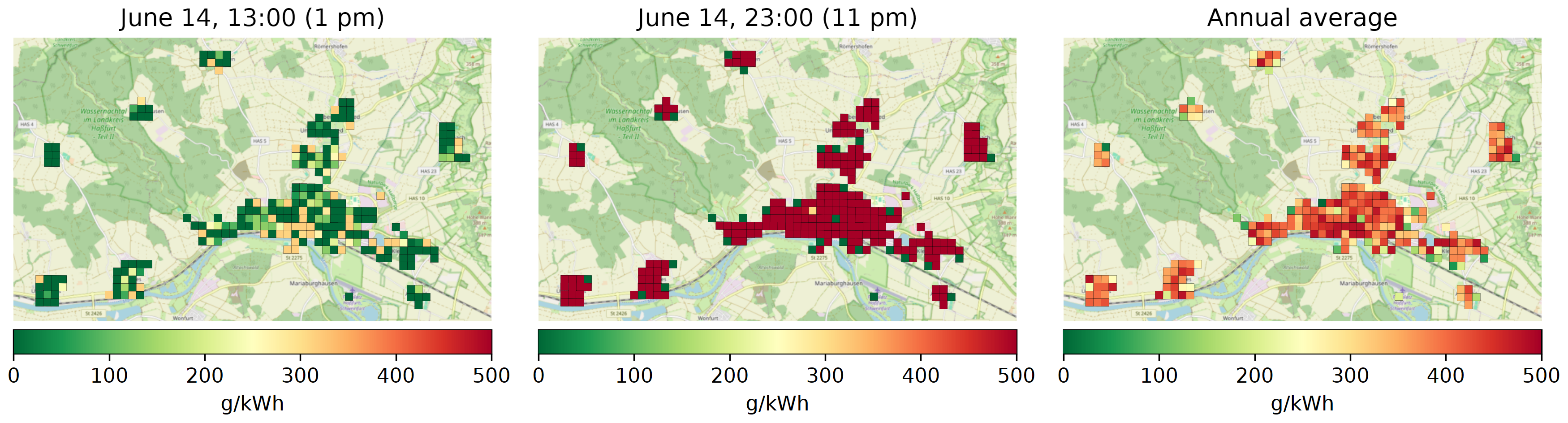}}
    \caption{Results of the local carbon intensity factors computation per 200m x 200m grid cell of the town of Ha{\ss}furt in Germany for the day with highest shortwave radiation in 2021 on the left. In the center, the results at where most cells require grid demand. The annually averaged carbon intensity factors are presented in the right plot.}
    \label{fig:results_of_carbon_intensity_factors}
\end{figure*}

\subsection{Local carbon intensity factors}
Figure~\ref{fig:results_of_carbon_intensity_factors} illustrates the spatially and temporally resolved carbon intensity factors computed for the grid cells.
The left plot displays the results for 1:00~PM on June 14\textsuperscript{th}, 2021, which corresponds to the day with the highest observed shortwave radiation throughout the year.
During this period of peak solar generation potential, many grid cells exhibit notably reduced carbon intensity values due to strong local \PV feed-in.
This is especially evident in residential areas with high rooftop \PV penetration.
Despite the widespread feed-in, considerable spatial variation remains: Over 50\% of all cells show a carbon intensity factor of zero—indicating full local coverage of electricity demand—but in areas with limited or no local \ac{PV}, intensities reach up to 316~g/kWh.
These differences are primarily driven by the uneven distribution of \PV systems across the town (see also Fig.~\ref{fig:pv_installations_per_cell}).

In contrast, the center plot of Fig.~\ref{fig:results_of_carbon_intensity_factors} shows the same area at the subsequent night, i.e., 11:00~PM on the same day, where most grid cells rely predominantly on imported grid electricity, which at this hour has a carbon intensity of 615~g/kWh.
This leads to a relatively homogeneous distribution of carbon intensity across the town, though some deviations remain.
Notably, a few cells benefit from local dispatchable generation, particularly \ac{CHP} plants, which reduce their carbon intensity significantly compared to their neighbors.
This highlights the important role of considering all local generation technologies beyond solar \PV for computing local carbon intensity factors.

The right plot of Fig.~\ref{fig:results_of_carbon_intensity_factors} presents the annual average carbon intensity per grid cell, revealing persistent spatial heterogeneity across the town.
Significant variations exist even between adjacent cells, underscoring the influence of localized generation capacities and potentially demand patterns.
Whereas 50\% of all cells show an average factor below 396~g/kWh, the maximum is notable higher with a value of 481~g/kWh.
Collectively, these examples demonstrate that local carbon intensity factors exhibits notable daily fluctuations, driven largely by renewable generation availability, and significant spatial variations, with the highest annual average intensities observed in the town center of Ha{\ss}furt.

\subsection{Estimation of local Carbon Intensity Factors}
\label{sec:results_estimation_of_factors}
The predictive models developed in this study are utilized for both intra-cell regression and inter-cell transferability. Intra-cell regression calculates the the carbon intensity of a cell based on the current solar radiation, time of day and the characteristics of the cell, where the model is trained on previous time frames. This approach is particularly useful when only a limited time frame of smart meter data is available and helps to reduce model complexity through the vastly reduced input requirements. Our testing has demonstrated a median $R^2$ score exceeding 0.92 across all cells and a median Relative Root Mean Squared of 0.09 for an XGBoost model. Therefore, a short period of smart meter data is sufficient to develop an effective surrogate model that transmission operators can use to estimate localized grid emissions.

Inter-cell transferability refers to the model's ability to predict carbon intensity in previously unseen cells, after being trained on data from other cells, possibly within the same cluster.
By training a generalized XGBoost model on all cells, excluding an arbitrary test set comprising 20\% of the cells, we achieve a RRMSE score of~1.23 on the test cells, which notably is worse then using country-wide grid emissions as a surrogate for local carbon emission intensity which leads to a RRMSE of~1.15. However, since the cells exhibit vastly different characteristics, we use the previously identified clusters and train a separate model for each cluster. For clusters with only one cell, we train the model on all cells except the outliers. Table \ref{PredictionResults} shows the RRMSE and the $R^2$-Score for each cluster and compares the three different prediction models, a decision trees, a XGBoost model and the TabNet neural network.

\begin{figure}[h!]
    \centering
    \includegraphics[width=0.7\linewidth]{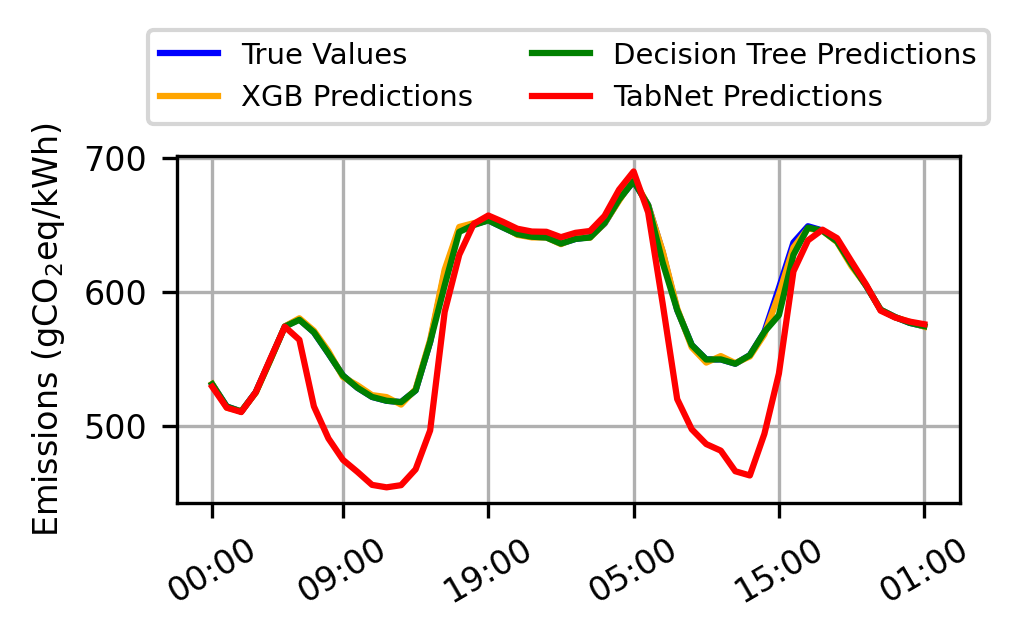}
    \caption{Comparison of Predictions for two consecutive days in March}
    \label{fig:preds}
\end{figure}

\begin{table}[h!]
    \caption{Prediction Metrics for XGBoost (XGB), Decision Trees (DT) and TabularNet Neural Network (TabNet)}
    \label{PredictionResults}
    \centering
    {
    \footnotesize
    \begin{tabular}{c|r|r|r|r|r|r}
                & \multicolumn{3}{c|}{$R^2$} & \multicolumn{3}{c}{RRMSE} \\
        Cluster &  DT & XGB & TabNet & DT  & XGB & TabNet \\\hline
        1 & 0.91 & 0.94 & 0.15 & 0.14 & 0.12 & 0.46 \\
        2 & 0.62 & 0.92 &  0.8 & 0.32 & 0.15 & 0.24 \\
        3* & 0.54 & 0.67 & -2322 & 0.67 & 0.56 & 48 \\
        4 & 0.80 & 0.88 & 0.79 & 0.24 & 0.18 & 0.25 \\
        5 & 0.96 & 0.98 & 0.92 & 0.07 & 0.05 & 0.1 \\
        6* & -0.19 & 0.74 & 0.72 & 1.12 & 0.52 & 0.53 \\
        7* & 0.39 & 0.74 & 0.04 & 0.59 & 0.39 & 0.73 \\
        8 & 0.82  & 0.86 & 0.82 & 0.18 & 0.16 & 0.18 \\
        9* & 0.89 & 0.86 & 0.89 & 0.14 & 0.16 & 0.14 \\
        
    \end{tabular}
    \\\vspace{5pt}
    \parbox{\linewidth}{
        \small $^*$ One cell clusters with highly irregular characteristics.
    }
	}
\end{table}
We see that XGBoost outperforms regular decision trees and the TabNet model for every cluster except cluster 9. Generally, a trend is visible that for the outlier clusters with only one cell, the general model trained on all other cells has a lower predictive quality. For one cluster each, the Decision Tree model (Cluster 6) and the TabNet model (Cluster 3) fail to produce any significant prediction capability. However, for the clusters with a higher number of cells, the predictive quality achieved is quite high with $R^2$ scores above 0.9 for the best majority. Negative $R^2$ scores occur for two clusters with only one cell where the model performs worse than using the average emissions as a predictor. This is most likely due to a combination to cell characteristics not observed in the training set which lead to a very poor performance of the predictor models.
In terms of the Relative Root Mean Squared Error, which captures the relative deviation of the prediction to the mean of the actual value, the XGB model is able to significantly outperform the baseline, country-level carbon intensity factors, which has a RRMSE of 1.15. Interestingly, the prediction performs well for both urban cells found in cluster 5 with a high density of buildings and rural cells represented in cluster 4 which represents suburban district characteristics. 
Fig. \ref{fig:preds} shows the prediction of the different models for an exemplary day of a previously unseen cell in cluster 5. It shows that the both the XGBoost and the Decision Tree model are able to accurately predict the carbon intensity of a previously unseen cell based on the cells characteristics and the current radiation and grid emission level. The neural network model on the other hand, captures the general trend but fails to outperform its decision tree based counterparts and exhibits a significantly longer training time. For this use case, it can therefore not be recommended.

The XGBoost model reveals significant variations in feature importance across clusters. While network emissions are the most influential feature in clusters 1, 2, and 5, total annual electricity demand dominates in cluster 4, and current global irradiation is most important in cluster~8. Generally, the top five features across most clusters include network emissions, global irradiation, installed PV power, total yearly electricity demand, and the number of installed wind turbines, although some clusters show deviations with other features occasionally ranking among the top five. This shows that even with little information about a cell, a highly accurate prediction can be made on the local carbon intensity through the use of a surrogate model when solar radiation and global grid emissions are known as exogenous input.

\subsection{Use Case}
To assess the potential benefits of considering local carbon emissions, we analyze an average building in the study region and simulate the integration of a large residential \ac{BESS} with a capacity of 20 kWh and a C-rate of 1. The storage capacity is chosen to cover the building’s daily electricity demand, allowing the battery to be charged from the grid during periods of low carbon intensity and discharged during periods of high intensity. A linear optimization model based on \cite{Schiller2024} is implemented to minimize total carbon emissions through optimal battery operation. When optimizing under the assumption of global emission factors, the resulting emissions are overestimated by 54 kgCO\textsubscript{2}eq or 9~\% compared to the actual emissions calculated with local carbon intensity. If the optimization were instead conducted directly with local emission data, total emissions could be further reduced by an additional 28 kgCO\textsubscript{2}eq.

\section{Discussion}
\label{sec:discussion}
As highlighted by \cite{2025_Sugano_SpatioTemporalCarbonIntensity}, relying solely on grid-average emission values is inadequate for accurate environmental assessments within urban regions. Our results confirm this, demonstrating that spatially-resolved carbon intensity information offers valuable insights.
Such granular information can support operational demand shifting and transparent, location-specific emission feedback.

Turning to the estimation or surrogate model approach (as detailed in Sec.~\ref{sec:results_estimation_of_factors}), the results indicate that predicting local carbon intensity at an hourly granularity is feasible for typical grid cells, even without access to detailed smart meter data for every consumer.
Cells characterized by a high \ac{RMSE} and a low coefficient of determination were typically those encompassing a small number of buildings. Potential mitigation strategies could involve adjusting the grid cell sizing or merging adjacent cells with low building counts to create more robust units.
\added{A major advantage of the surrogate model is that it enables fast inference without requiring a full physical simulation as described in Sec.~\ref{sec:method_carbon_intensity_factor_computation}, making it suitable for real-time operational use.}

\subsection{Limitations}
The proposed methodology for computing carbon intensity factors using smart meter data (outlined in Sec.~\ref{sec:method_carbon_intensity_factor_computation}) is inherently dependent on the availability and quality of external datasets.
This includes meteorological data, specifically solar radiation measurements, and the accuracy of meta data for \PV modeling.
Inaccuracies or missing data in these inputs will inevitably propagate errors into the calculated carbon intensity factors.
Moreover, the utilized \ac{PV} simulation model should be calibrated with real data, as we have encountered for the well-known PVlib.
\added{Finally, while the surrogate model is designed to rely on features typically available to grid operators and municipalities, its quantitative transferability to other geographical regions has not yet been empirically validated due to a lack of the data required to compute local carbon intensity factors as detailed in Sec.~\ref{sec:method_carbon_intensity_factor_computation} for other regions.}

\section{Conclusion}
\label{sec:conclusion}
Local carbon emission factors can vary strongly across different grid cells within the same town or city.
Our work presents a method for the computation of local carbon intensity factors based on smart meter data and additional meta data.
Therefore, a town was used that has very rural parts on the one hand, but also more metropolitan districts on the other.
Building upon this, a step towards transferability is presented by describing a method for estimating these factors based on a small set of input parameters that can be obtained quicker than smart meter data for all buildings within a town.
As a result, this study underscores the significant spatial and temporal variations in local carbon intensity across different grid cells within a town.
Moreover, the results show that all local generation technologies should be considered when computing these local intensities.
The results of this paper, particularly the surrogate model, can be used to foster the reduction of carbon emissions, for example by enabling informed shifting of \ac{EV} charging processes or data center workloads to regions with lower carbon intensity.
In future work, the proposed method should be applied to additional countries with diverse settlement structures to develop a surrogate model that is applicable in every region of our world.

\section*{Acknowledgment}
A special thanks goes to the distribution system operator of Ha{\ss}furt, Stadtwerk Ha{\ss}furt GmbH, for supporting us with data.

\bibliographystyle{IEEEtran}
\bibliography{referencesP12}

\end{document}